\documentclass{article}
\usepackage{graphicx}
\usepackage{amsmath,amssymb}
\usepackage{graphics,color,array,calc,rotating,epsfig,psfrag}
\numberwithin{equation}{section}
\usepackage{cite}
\usepackage{bm}
\usepackage{dcolumn}
\usepackage{float}
\usepackage{subfigure}
\usepackage{amsfonts}
\usepackage[mathscr]{eucal}
\def\be{\begin{equation}} \def\ee{\end{equation}}
\def\bea{\begin{eqnarray}} \def\eea{\end{eqnarray}}

\newcommand{\nn}{\nonumber}
\begin{document}
\baselineskip 18pt%
\begin{titlepage}
\vspace*{1mm}%
\hfill%
\vspace*{15mm}%
\hfill
\vbox{
    \halign{#\hfil         \cr
          } 
      }  
\vspace*{20mm}

\begin{center}
{\large {\bf   Gravitational measurements in higher dimensions
}}\\
\vspace*{5mm}
{  Davood Mahdavian Yekta\footnote{d.mahdavian@hsu.ac.ir}, S. A. Alavi\footnote{s.alavi@hsu.ac.ir}, Majid Karimabadi}\\

\vspace*{0.2cm}
{$^{}$ Department of Physics, Hakim Sabzevari University, P.O. Box 397, Sabzevar, Iran}\\
\vspace*{1cm}
\end{center}
\begin{abstract}

We attempt to study three significant tests of general relativity in higher dimensions both in commutative and non-commutative spaces. In the context of non-commutative geometry, we will consider a solution of the Einstein’s equation in higher dimensions with a source given by a static, spherically symmetric Gaussian distribution of mass. The resulting metric would describe a regular or curvature singularity free black hole in higher dimensions. The metric should smoothly interpolate between Schwarzschild geometry at large distance and de-Sitter spacetime at short distance. We will consider gravitational redshift, lensing, and time delay in each sector. It will be shown that, compared to the four-dimensional spacetime, there can be significant modifications due to the presence of extra dimensions and the non-commutative corrected black holes. Finally, we shall attempt to obtain a lower bound on the size of the extra dimensions and on the mass needed to form a black hole in different dimensions.
\end{abstract}

\end{titlepage}
%
\section{Introduction}

Since Einstein proposed his general theory of relativity in 1915, a lot of research has been devoted to unify General Relativity (GR) and Electromagnetism as two fundamental interactions in nature. However, the early proposals date back to the 1920s, through Kaluza–Klein theory to unify these interactions  \cite{Kaluza:1921tu,Klein}, that was a classical unified field theory built in five-dimensional spacetime. Recently, motivated by string theory as a requirement for describing a consistent theory of quantum gravity, extra dimensions have been the subject of much attention. Besides string theory, there are some other theories proposing the necessity of extra dimensions:
\begin{itemize}
\item Large extra dimensions, mostly motivated by the ADD model, by Arkani-Hamed, Dimopoulos, and Dvali together with Antoniadis in Refs. \cite{ArkaniHamed:1998rs,Antoniadis:1998ig} to solve the hierarchy problem in which the difference between the Standard Model interactions and GR manifests itself notably in their dissimilar coupling strengths. While the electromagnetic, weak and strong forces differ by just six orders of magnitude, the gravitational interaction falls apart by further thirty-three orders.
\item Warped extra dimensions, such as those proposed by the Randall–Sundrum (RS) model \cite{Randall:1999ee}, in which our observable universe is modeled as a four dimensional hyper surface, known as the 3-brane, embedded in a five dimensional space, usually  called the bulk. The novel idea of the Brane world is that all the gauge interactions, described by the Standard Model, are confined to live in the 3-brane while the gravitational interaction can spread into the fifth dimension of the space.
\item Universal extra dimensions, proposed and first studied in Ref. \cite{Appelquist:2000nn}, assume, at variance with the ADD and RS approaches, that all fields propagate universally in the extra dimensions.
\end{itemize}
The size and the shape of extra dimensions should be related to the fundamental energy scales of particle physics: the cosmological scale, the density of dark energy, the TeV electroweak scale, or the scale of ultimate unification. More likely, the extra dimensions are microscopic, in this case high-energy particle accelerators \cite{Giddings:2001bu,Dimopoulos:2001qe} and cosmic-ray experiments \cite{Nagano:2000ve,Emparan:2001kf} are the only means to detect their physical effects. The LHC experiments will have direct sensitivity to probe extra dimensions, through the production of new particles that move in the extra space. There is also a chance that, due to the existence of extra dimensions, microscopic black holes may be detected at the LHC \cite{Dimopoulos:2001hw,Cheung:2001ue} or in the highest energy cosmic rays \cite{Feng:2001ib,Anchordoqui:2001cg}.

On the other hand, Einstein’s work derived gravitation from the underlying spacetime concept and was not provoked by observational facts but was motivated on a purely theoretical basis, while this theory  fundamentally has changed our understanding of spacetime, mass, energy, and gravity. GR had some features and implications further than  Newton’s theory of gravitation, namely, light bending, time dilation, and gravitational redshift  \cite{Weinberg}.These effects have been verified experimentally and to this date are being tested to higher and higher accuracies. The gravitational waves which were recently detected by LIGO and Virgo collaborations  \cite{Abbott:2016blz}are also another profound implication of GR. The detected signals perfectly agree with predictions based on black holes in GR up to  $5\sigma$ \cite{TheLIGOScientific:2016pea}.

Gravitational redshift is a very useful tool in astrophysics. This phenomenon has been confirmed by Pound–Rebka experiment in 1959 \cite{Pound:1960zz}.It helps us test our knowledge about the structure of those stars whose internal structures are different from the Sun and other normal stars.
Gravitational lensing occurs when light rays pass close to a massive body and it was confirmed by Eddington for the first time in 1919\cite{Dyson:1920cwa}.
About one century after the first measurement, gravitational lensing is still one of the major tools of cosmology\cite{Peebles:2004qg,Song:2008xd}, astrophysics\cite{Fomalont:1976zz,Fomalont:2009zg} and astronomy \cite{Bull:2015lja,Rusin:2002tq,Reyes:2010tr}.

Time dilation measures the amount of time elapsed between two events by observers situated at different distances from a gravitational mass. The light travel time delay is sometimes called the fourth classical test of GR and was first introduced by Shapiro in 1964  \cite{Shapiro:1964uw}. A significant improvement was reported in \cite{Bertotti:2003rm} from Doppler tracking the Cassini spacecraft on its way to Saturn. In addition to the above theoretical motivation(s), there are advances in technologies concerning the high precision measurement of time and frequency  namely, optical lattice clocks \cite{Amenomori:2006bx} and auto-seconds laser technologies \cite{Takamoto:2005}.
Time delay corrections are also very important in Global Positioning Systems (GPS) \cite{Ashby:2003vja}. The clocks on GPS satellites tick faster than the clocks on earth’s surface, so we have to place a correction into the satellite measurements.

In addition to the idea of extra dimensions, the other important implication motivated by string theory is the non-commutativity of space \cite{Seiberg:1999vs,Ardalan:1998ce,Kempf:1994su,AmelinoCamelia:2008qg}. It has drawn a lot of interest in a wide range of areas from condensed matter physics to cosmology, high energy physics, and astrophysics \cite{Connes:1997cr,Fatollahi:2006tp,Zhang:2011zzg}.
The simplest non-commutativity that one can postulate is the commutation relation $[x_{i},x_{j}]=i \theta_{ij}$, where $\theta ij$ is an antisymmetric (constant) tensor of dimension $(length)^2$. The parameter $\theta$ measures the amount of coordinate non-commutativity in the coordinate coherent states (CCS) approach \cite{Snyder:1946qz,Doplicher:1994tu} in which the concept of point-like particle becomes physically meaningless and must be replaced with its best approximation, i.e., a minimal width Gaussian distribution of mass.

In fact, the CCS approach to non-commutative effects can cure the singularity problems at the final stage of the black hole evaporation. This effective approach may be considered as an improvement to semi-classical gravity and a way to understand the non-commutative
effects. Motivated by this idea, the Schwarzschild black holes inspired by non-commutative geometry studied in \cite{Nicolini:2005vd}, was extended to the Reisnner-Nordstrom model in \cite{Ansoldi:2006vg,Alavi:2009tn}, and has been generalized to higher dimensions in \cite{Rizzo:2006zb}, and  charged black holes in higher dimensions \cite{Nozari:2007ya,Spallucci:2009zz,Gingrich:2010ed}.Furthermore, in recent years we have witnessed a significant interest in this non-commutative approach from cosmology  \cite{Calmet:2005mc,Alavi:2004aq}, holography \cite{Pramanik:2014mya,Zeng:2014xpa,Pramanik:2015eka} and the black hole physics \cite{Nicolini:2009gw,Liang:2012vx,Ghosh:2017odw,Wei:2015dua,Myung:2006mz,Nozari:2007rh,Banerjee:2008gc,Miao:2015npc}.

On the other hand, studying the effects predicted by GR are important in the vicinity of compact objects, such as neutron stars and black holes. The first light detected from regions close to the black holes was discovered by the ASCA satellite \cite{Tanaka:1995en,Yaqoob:1998ef}. In the case of astrophysical effects in the vicinity of higher dimensional black holes one can see \cite{Connell:2008ek} and references therein.   The effects of such black hole parameters as charge and rotation on the gravitational lensing have been studied in \cite{Kraniotis:2005zm} for Kerr metric and in \cite{Briet:2008mz} for charged solutions (from free charges like RN black holes to geometrical charges like Kaluza-Klein black holes coming from compactification of extra dimensions). It has also been investigated to determine how the detection of lensed images of black holes determine the form of the black hole metric in \cite{BinNun:2010se}, but the calculations in \cite{Kraniotis:2005zm,Briet:2008mz,BinNun:2010se} are in the strong-field regime of GR.
The authors in \cite{Pardo:2018ipy} have shown that the Virgo and LIGO results for GW170817 data \cite{Abbott:2016blz} have the best consistency with GR but their results do not hold for extra dimensional theories with compact extra dimensions in strong energy limit \cite{ArkaniHamed:1998rs,Antoniadis:1998ig,Randall:1999ee} and for theories with larger extra dimensions, typically cosmological distances \cite{Dvali:2000hr}, in the weak-field regime.

The purpose of the current work is to obtain explicit expressions for the three aforementioned GR effects in the gravitational field of a black hole in commutative and non-commutative spaces with extra dimensions. Inspired by this idea, we do investigate deviations from GR predictions due to the gravitational leakage into the extra dimensions, and we are more interested in some possible observational or experimental consequences of extra dimensions in such gravitational systems. This issue deserves further research along the lines that we have already proposed in \cite{Karimabadi:2018sev}.

 The structure of the paper is as follows: Section II, introduces the Schwarzschild black hole in higher dimensions, in some detail. Similarities and differences of GR measurements in four and extra dimensions are illustrated in Section III. In particular, there are three figures which make the comparison easier and clearer. As our results show, if spacetime is truly a higher dimensional space then its implications should appear in gravitational measurements around black holes.

The goal of section IV, is to study the effects of the non-commutativity of space on higher dimensional GR measurements. We first present some preliminaries of black holes in non-commutative higher dimensional spaces. Then, we obtain a minimum mass to form a black hole in each extra dimension and investigate the gravitational measurements in non-commutative higher dimensional spaces. Although theories with extra dimensions has received much attention in recent years, unfortunately the size of extra dimensions has not been investigated properly. To tackle this problem, we finally obtain lower bound on the size of extra dimensions using the bound obtained for the non-commutative length scale in our previous work \cite{Karimabadi:2018sev}.

\section{Schwarzschild black hole in higher dimensions}

Amongst the various types of black hole solutions of Einstein field equations, a natural higher dimensional generalization of the Schwarzschild metric, also known as the Schwarzschild-Tangherlini metric \cite{Tangherlini:1963bw}, has been assumed to be stable, like its four-dimensional counterpart. The spacetime around such an uncharged, stationary, spherically symmetric black hole in $(d+1)$ dimensions is described by
\be\label{Sch1} ds^2=B(r) dt^2-B^{-1}(r) dr^2-r^2 d\Omega^2_{d-1}\,,\ee
where $d\Omega^2_{d-1}$ denotes the element of unit $(d-1)$-sphere with area $A_{d-1}=\frac{2 \pi^{d/2}}{\Gamma(d/2)}$ and  $B(r)$ is given by
\be\label{func1}B(r)=1-\frac{\mu_0}{r^{d-2}}\,.\ee

The constant parameter $\mu_0$ is related to the mass of the black hole by \cite{Myers:1999psa}
\be\label{sm} M=\frac{(d-1) A_{d-1}\mu_0}{16 \pi G_{d+1}} \,,\ee
where $G_{d+1}=G_4 L^{d-3}$ is the $(d+1)$-dimensional gravitational constant and $L$ is the size of the extra dimensions, so
\be\label{func2}B(r)=1-\frac{8 MG_4 L^{d-3}\, \Gamma[\frac{d}{2}]}{(d-1)\,\pi^{\frac{d}{2}-1} \, r^{d-2}}\,.\ee
For later convenience, we use $G_4=1$ and define dimensionless variables $x=\frac{r}{\ell_{p}}$, $\eta=\frac{M}{\ell_{p}}$, and $\alpha=\frac{L}{\ell_{p}}$ where $\ell_{p}$ is the Planck's length.

It is worth mentioning that if a gravitational radius of a black hole is much smaller than the characteristic length of the extra dimensions, then the black hole can be very well described by asymptotically flat solutions like those in \cite{Tangherlini:1963bw,Myers:1986un,Scardigli:2003kr,Nouicer:2007pu} for higher dimensional static and rotating black holes (a review on higher dimensional black holes can be found in \cite{Emparan:2008eg}, see also references therein). Thus, from this point of view, one might conclude that there should not be any practical difference between these kinds of black holes and the small black holes in the ADD and RS models given in the introduction.

\begin{figure}[H]
\centering
\includegraphics[width=10cm,height=6.5cm]{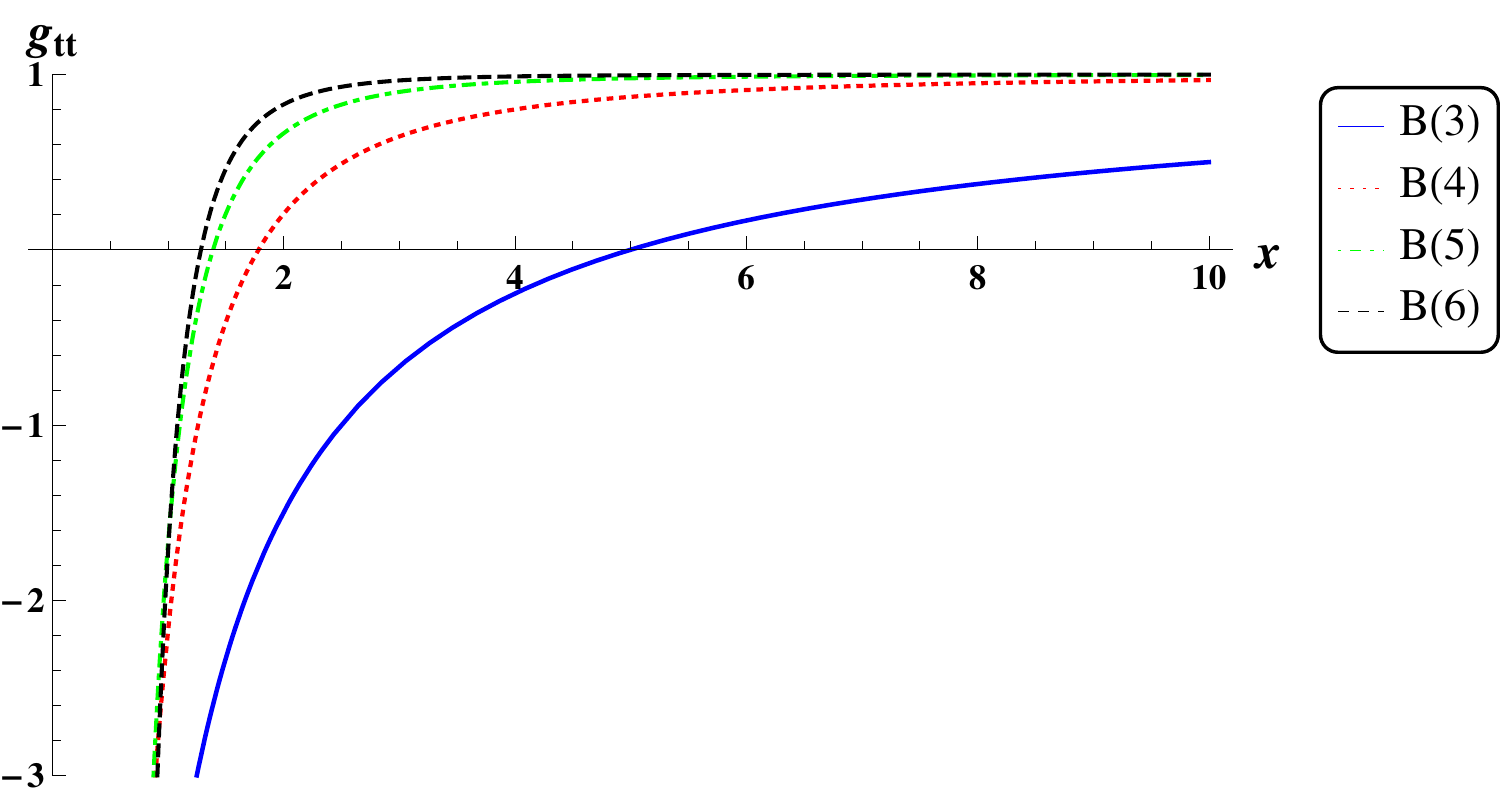}
\caption{The $g_{tt}$ component for $\alpha\!=\!1.5$ and $\eta\!=\!2.5$. The number in parentheses is the spatial dimension $d$, with this extra explanation that in units where $c=G=\hbar=1$, these values for $\alpha$ and $\eta$ are satisfactory.}
\label{f1}
\end{figure}

We have plotted the $g_{00}$ component of higher dimensional Schwarzschild metric (\ref{Sch1}) as a function of $x$ for different spatial dimensions in Fig.~(\ref{f1}). The location of the event horizon is determined by the equation $B(r)=0$, so as seen in Fig.~(\ref{f1}) this occurs in smaller distances in higher dimensions which asserts that gravity in four-dimensional spacetime is stronger than higher dimensions. This fact can also be checked by noting that, in higher dimensions the $g_{tt}$ curves tend more rapidly to the $g_{tt}$ of flat  spacetime.
\section{Gravitational effects in higher dimensions}
In this section, we are going to obtain expressions for the three aforementioned effects of GR in the case of an extra dimensional Schwarzschild black hole as the gravitational system. In order to compare the behaviour of extra dimensions with GR, we perform a numerical analysis by plotting the quantities. A general remark is in order here. The details of the calculations in this section for four dimensions could be found in Ref.~\cite{Weinberg}, and the calculations for extra dimensions could be carried out along the same lines.

{\bf{\emph{Redshift:}}} When the light passes in the opposite direction of a gravitational field, some of its energy is wasted and it is transmitted to redshift wavelength. The gravitational redshift is denoted by $z =\Delta \lambda/\lambda$ where $\Delta\lambda$ is the difference between the observed and emitted wavelengths and $\lambda$ is the wavelength of the source. However, for radiation emitted in a strong gravitational field, as that coming from the surface of a neutron star or close to the event horizon of a black hole, the gravitational redshift can be very large. Hence, there is a shift in the spectral lines of light around a Schwarzschild metric (\ref{Sch1}) which is given by the following maximum value
\be \label{reds1} z=\frac{\omega_1}{\omega_2}\Big|_{max}-1=\sqrt{\frac{B(r_2)}{B(r_1)}}-1\,,\ee
where $\omega_2$ and $\omega_1$ are the frequencies received by the observer and emitted by the source, respectively. (for more details see $\S\,14.3$ in Ref.~\cite{Weinberg}) When the light is emitted from radius $r_1$ and received at $r_2\rightarrow\infty$, then the redshift measured by an asymptotic observer turns out to be
\be\label{reds2} z=\left[1-\frac{8 M L^{d-3}\, \Gamma[\frac{d}{2}]}{(d-1)\,\pi^{\frac{d}{2}-1} \, r_1^{d-2}}\right]^{-1/2}-1.\ee

\begin{figure}
\centering
\includegraphics[width=12cm,height=7cm]{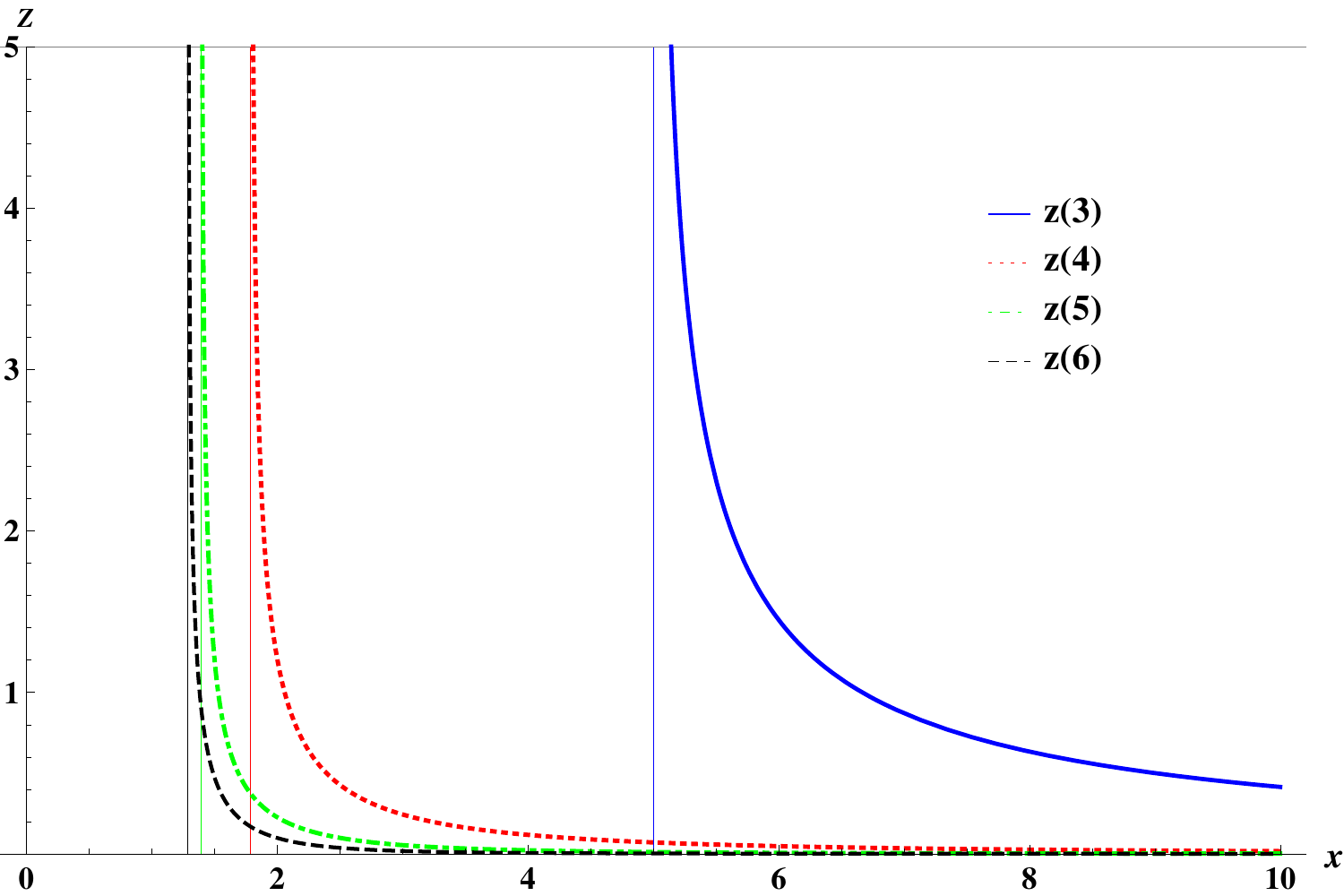}
\caption{Redshift for different values of $d$ in terms of $x$ for $\alpha=1.5$ and $\eta=2.5$. The vertical line shows the location of event horizon in each dimension and $x=r_1/\ell_p$.}
\label{f2}
\end{figure}

We have plotted the redshift factor (\ref{reds2}) for different spatial dimensions in Fig.~(\ref{f2}). Comparing the graphs confirms the statement that in higher dimensions the spacetime foam has lower curvature than GR or the gravity is diluted in extra dimensions. In other words, it can be inferred from the figure that the rate of increase in redshift occurs at higher dimensions closer to the black hole event horizon.

\begin{figure}[H]
\centering
\includegraphics[width=10cm,height=4cm]{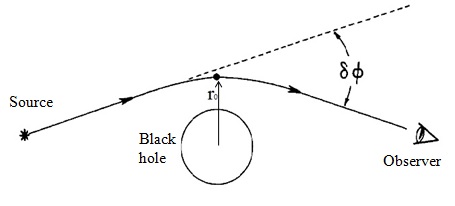}
\caption{Gravitational deflection of light around a massive object.}
\label{f10}
\end{figure}

{\bf{\emph{Deflection of light:}}} When the light passes close to a massive object such as a supernova or a black hole, it is deflected from its straight path by the value
\be\label{def} \Delta\phi=2 \int^{\infty}_{r_{\circ}} \frac{1}{r\sqrt{B(r)}} \,\left(\frac{r^2}{r_{\circ}^2}\frac{B(r_{\circ})}{B(r)}-1\right)^{-\frac12} dr-\pi\,,\ee
where $r_{\circ}$ is the closest distance to the massive object depicted in Fig.~(\ref{f10}). The details of the calculation can be found in $\S\,8.5$ of Ref.~\cite{Weinberg}.
The bending and delay of photons by the curvature of spacetime produced by a mass are proportional to $\gamma+1$ ( $\gamma$ is called the parameterized post-Newtonian parameter), where $\gamma$ is one in GR but zero in the Newtonian
theory \cite{Weinberg}. Henceforth, we consider $\gamma=1$ and also ignore the modification of the Heisenberg uncertainty principle to a Generalized Uncertainty Principle (GUP) given in \cite{Kempf:1994su,Scardigli:2014qka}.

\begin{figure}
\centering
{\includegraphics[width=.5\textwidth]{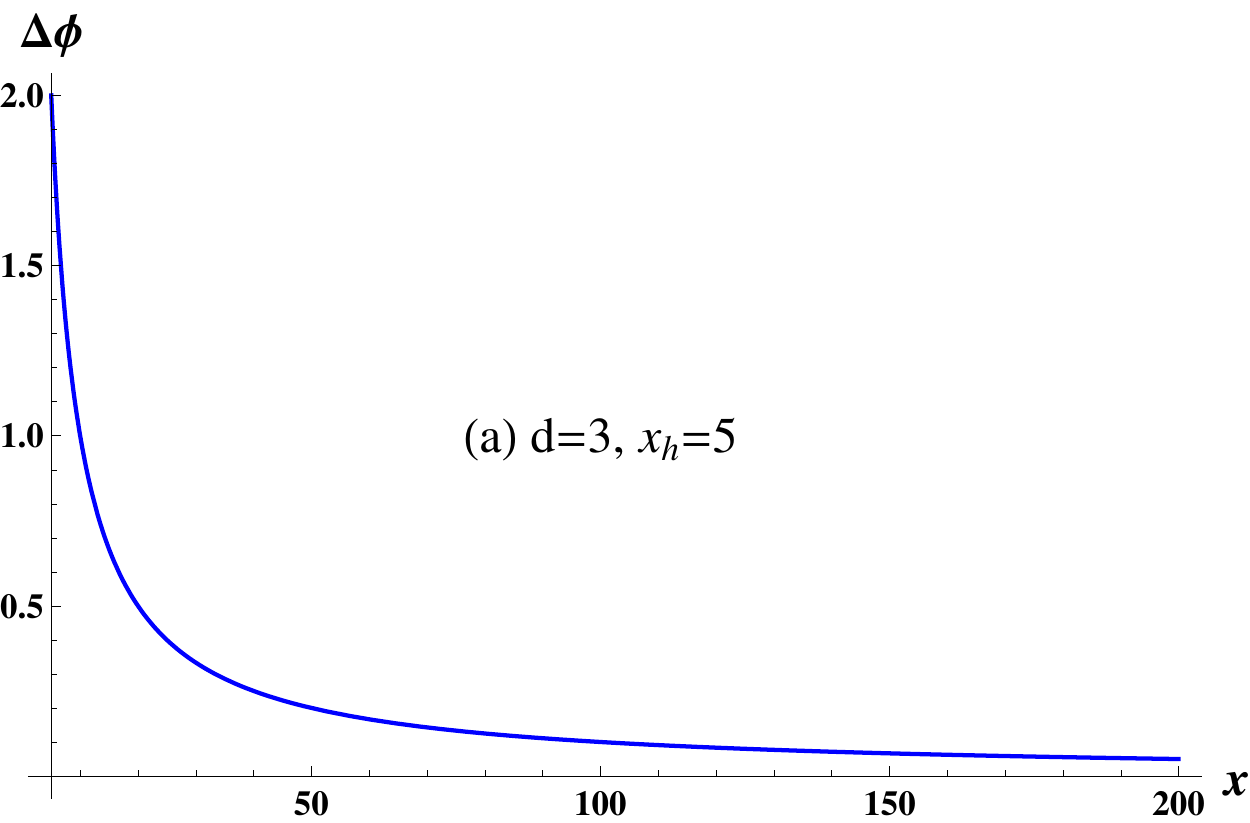}}
{\includegraphics[width=.5\textwidth]{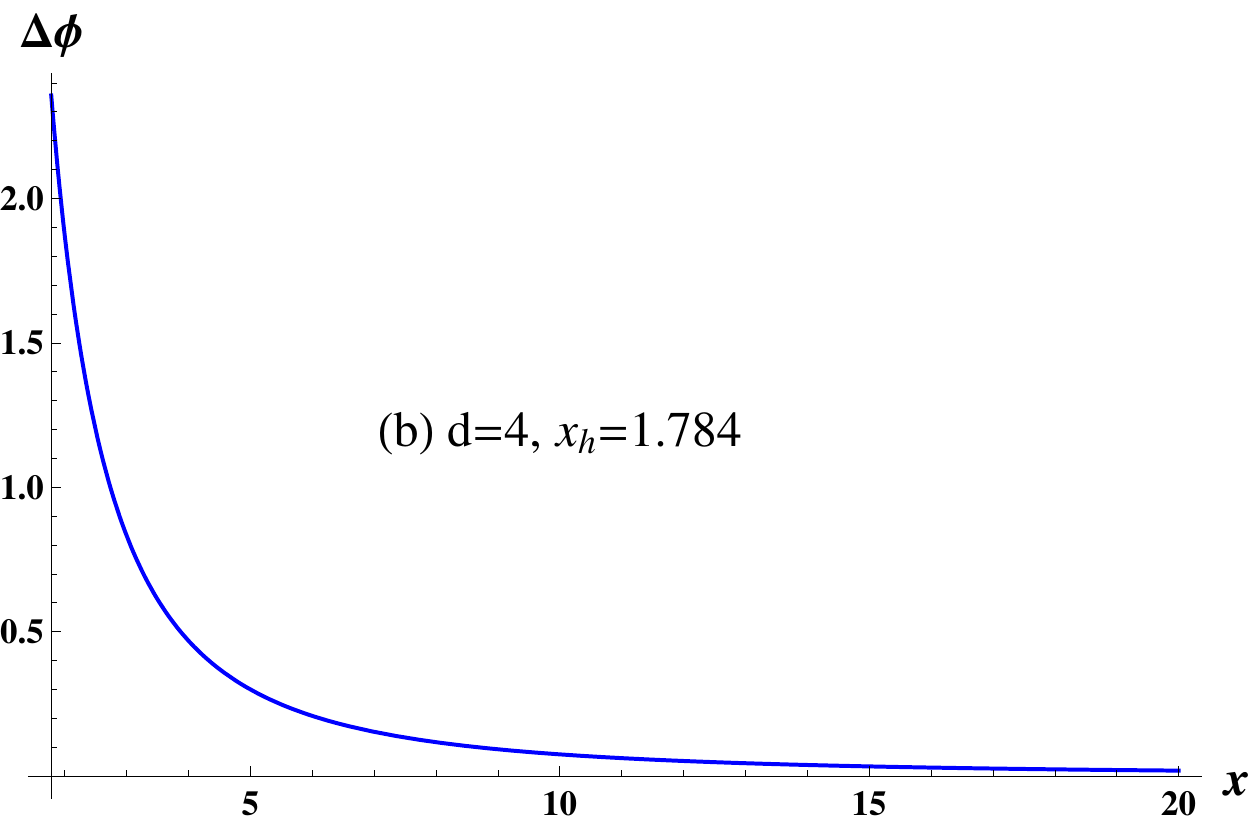}}
{\includegraphics[width=.5\textwidth]{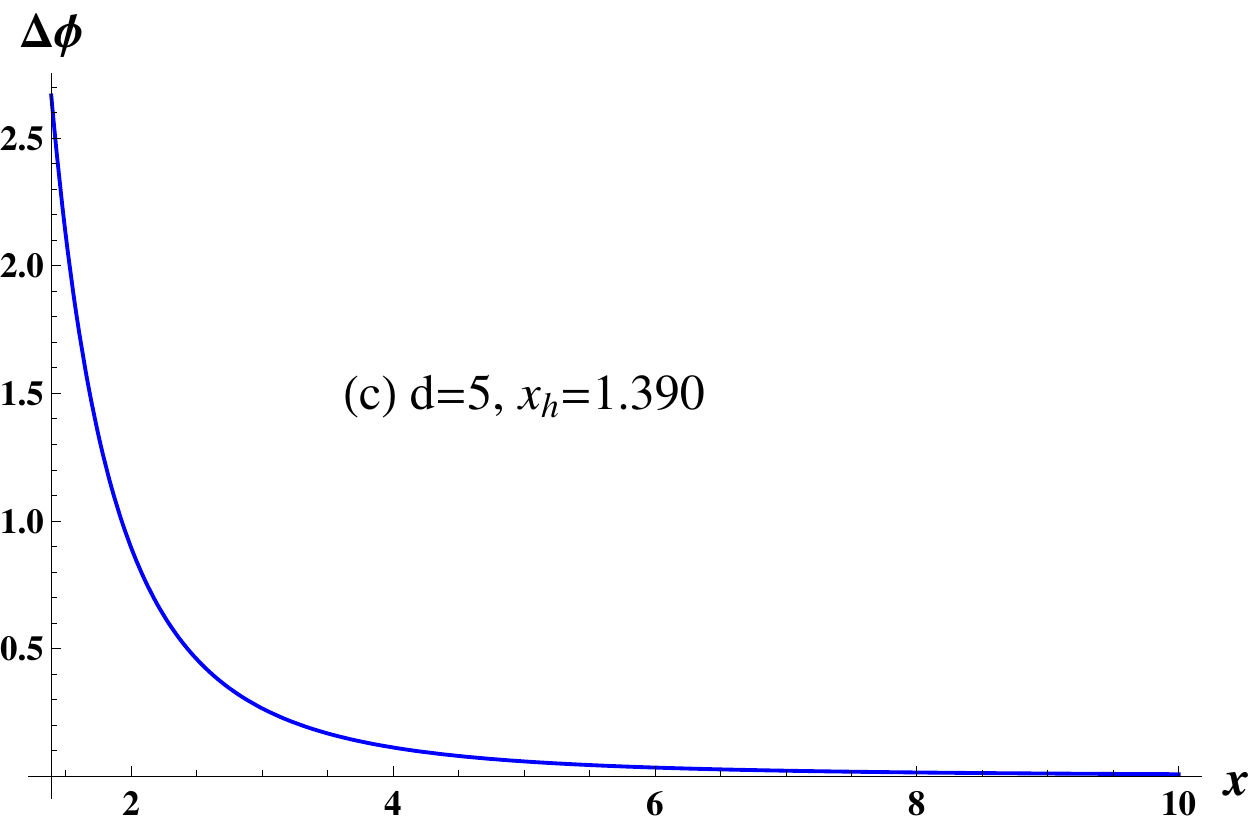}}
{\includegraphics[width=.5\textwidth]{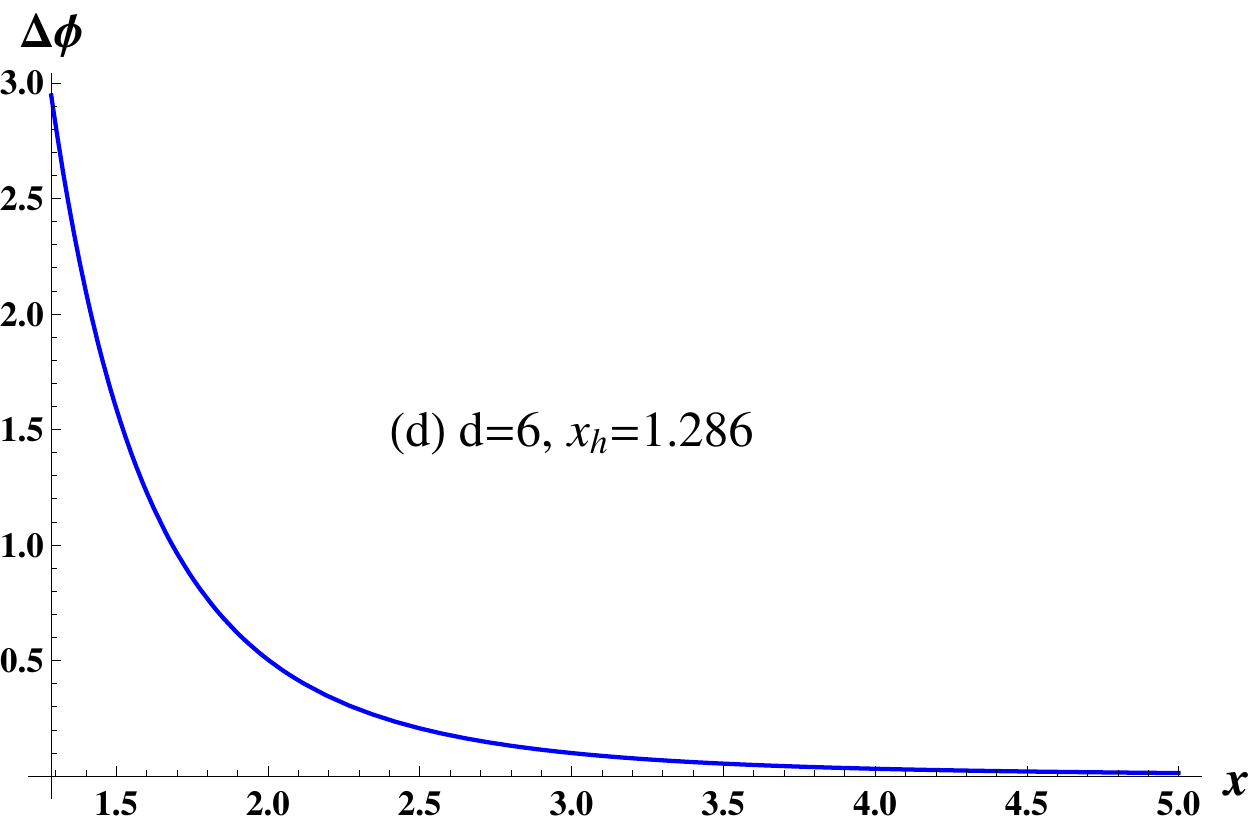}}
	    \caption{\small{Deflection of light around a Schwarzschild black hole for $\alpha=1.5$ and $\eta=2.5$ in diffenet spacetime dimensions.}}
\label{f3}
\end{figure}

The integration yields the following expression for bending of light in the vicinity of a Schwarzschild metric (\ref{Sch1}),
\be\label{Schdef} \Delta\phi=\frac{4 M L^{d-3} \, \Gamma[\frac{d-1}{2}]}{ \pi^{\frac{d-3}{2}}\,r_\circ^{d-2}}.\ee
The behaviour of this quantity has been plotted versus $r_\circ$ for different $d$ in Figs.~(\ref{f3}). We have used the dimensionless convention introduced in the previous section. The figures are plotted from the event horizon in each dimension. By comparing the figures, it is observed that as the dimension of spacetime increases, the deflection occurs at closer distances to the event horizon, or equivalently, the suppression takes place in short distances. Also, the maximum value of $\Delta \phi$, which happens in the horizon, increases by $d$. However, the general behaviour of the plots are the same for all dimensions. Thus, the results show that the deflection of light in higher dimensions is weaker than GR.

\vspace{1cm}
{\bf{\emph{Time delay:}}} According to GR massive objects curve the spacetime geometry, so the motion of different particles, such as photons, is affected by this curvature. Bending the spacetime causes the light path become longer than the straight path, therefore it takes more times to travel, consequently generates a time delay. The maximum round-trip excess time delay around the black hole described by (\ref{Sch1}) is given by \cite{Weinberg}
\be\label{td1} (\Delta t)_{max}=2\left[t(r_0,r_1)+t(r_0,r_2)-\sqrt{r_1^2-r_0^2}-\sqrt{r_2^2-r_0^2}\,\right],\ee
where the time required for light to go from $r_0$ to $r$ is
\be \label{td10} t(r_0,r)=\int_{r_0}^{r}\frac{1}{B(r)} \left(1-\frac{B(r)}{B(r_0)}\,\frac{r_0^2}{r^2}\right)^{\frac12}\,dr .\ee
The reader can see the details in $\S\,8.7$ of Ref.~\cite{Weinberg}.

\begin{figure}[H]
\centering
\includegraphics[width=6cm,height=4cm]{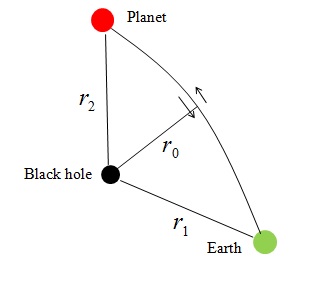}
\caption{The actual path of the radar reflection of photons from the earth to a planet and back.}
\label{f40}
\end{figure}

In order to calculate the excess (\ref{td1}), we first use the Robertson expansion for the integrand in (\ref{td10}). The leading terms in the expansion,i.e., $\sqrt{ r_1^2 - r_0^2}$ and $\sqrt{ r_2^2 - r_0^2}$ which are what we should expect if light traveled in straight lines at unit velocity, are canceled and only remain the dominant terms given by
\bea\label{tds} \Delta t &\!\!\!\!\simeq \!\!\!\!&\frac{4\eta}{\pi^{\frac{d}{2}-1} \alpha^{d-3}}\Bigg[ \frac{(d-2) \pi^{\frac32} i \csc [\frac{\pi d}{2}]}{x^{d-3} \Gamma \left(\frac{5-d}{2}\right)}-\frac{4}{(d-3)(d-1)}\nn\\
&&\left((\frac{\delta^3}{\delta^d}+\frac{\sigma^3}{\sigma^d})\Gamma \left(\frac{d}{2}\right)+\frac{6(d-3)-4(d-4)}{(d-1) x^{d-3}} (\frac{i}{2})^{d-1} \Gamma \left(\frac{4-d}{2}\right) \Gamma \left(d\right) \right)\Bigg],
\eea
where we have used the previous dimensionless variables $\alpha,\,\eta$, and $x$.
In fact, these terms evidently produce a GR delay in the time it takes a radar signal to travel to a planet and back. This excess delay is a maximum when planet is at superior conjunction and the radar signal just grazes the black hole; in this case $r_0$ is almost equal to the event horizon radius and is much smaller than the distances $r_1$ and $r_2$ of the black hole from the earth and the planet, respectively (see Fig.~\ref{f40}). For $d=3$, the excess time delay (\ref{tds}) gives the known result
\be\label{tds1} (\Delta t)_{max}\simeq 4\eta \left\{ 1+\ln{[\frac{4\delta\sigma}{x^2}]}\right\}. \ee
The dimensionless parameters $\delta=\frac{r_1}{\ell_p}$ and $\sigma=\frac{r_2}{\ell_p}$ are the orbital radius of the earth and of the reflecting planet around the center of the black hole illustrated in Fig.~\ref{f40}.

\begin{figure}
\centering
{\includegraphics[width=.5\textwidth]{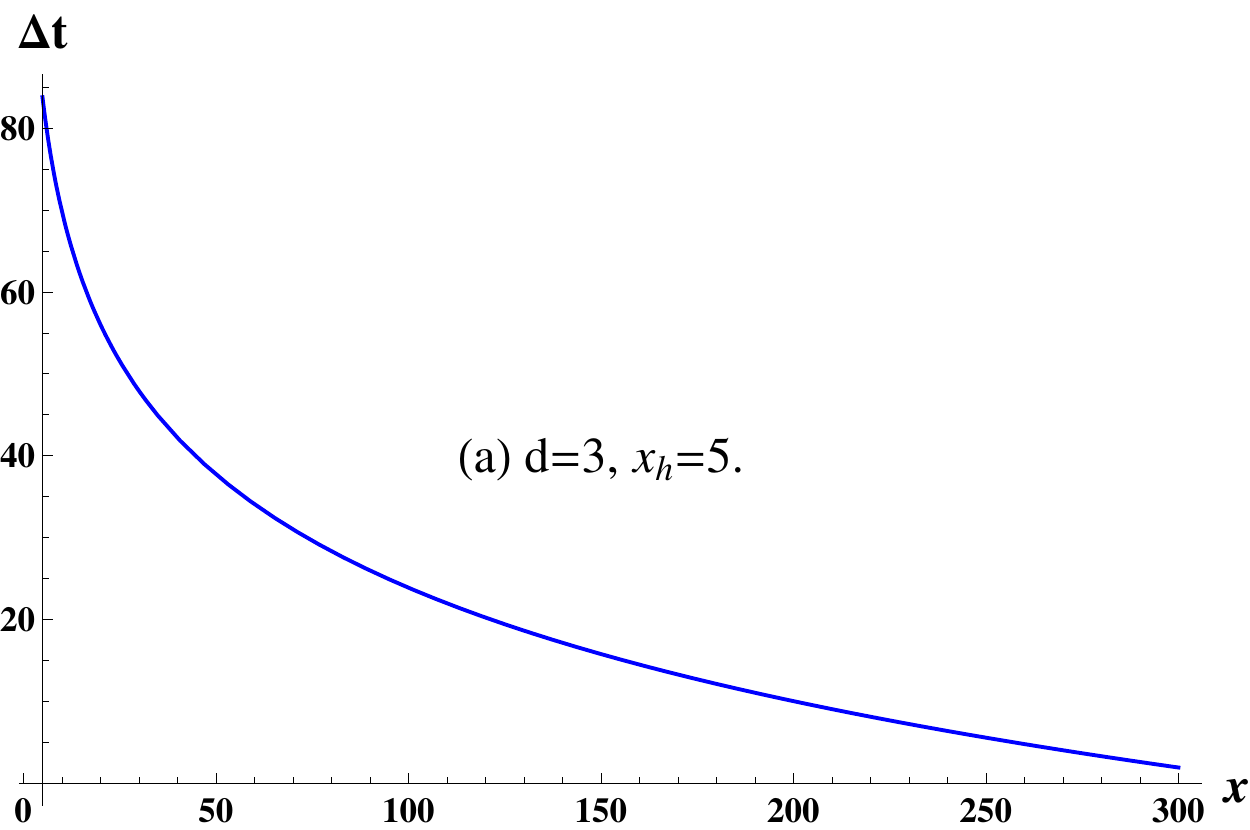}}
{\includegraphics[width=.5\textwidth]{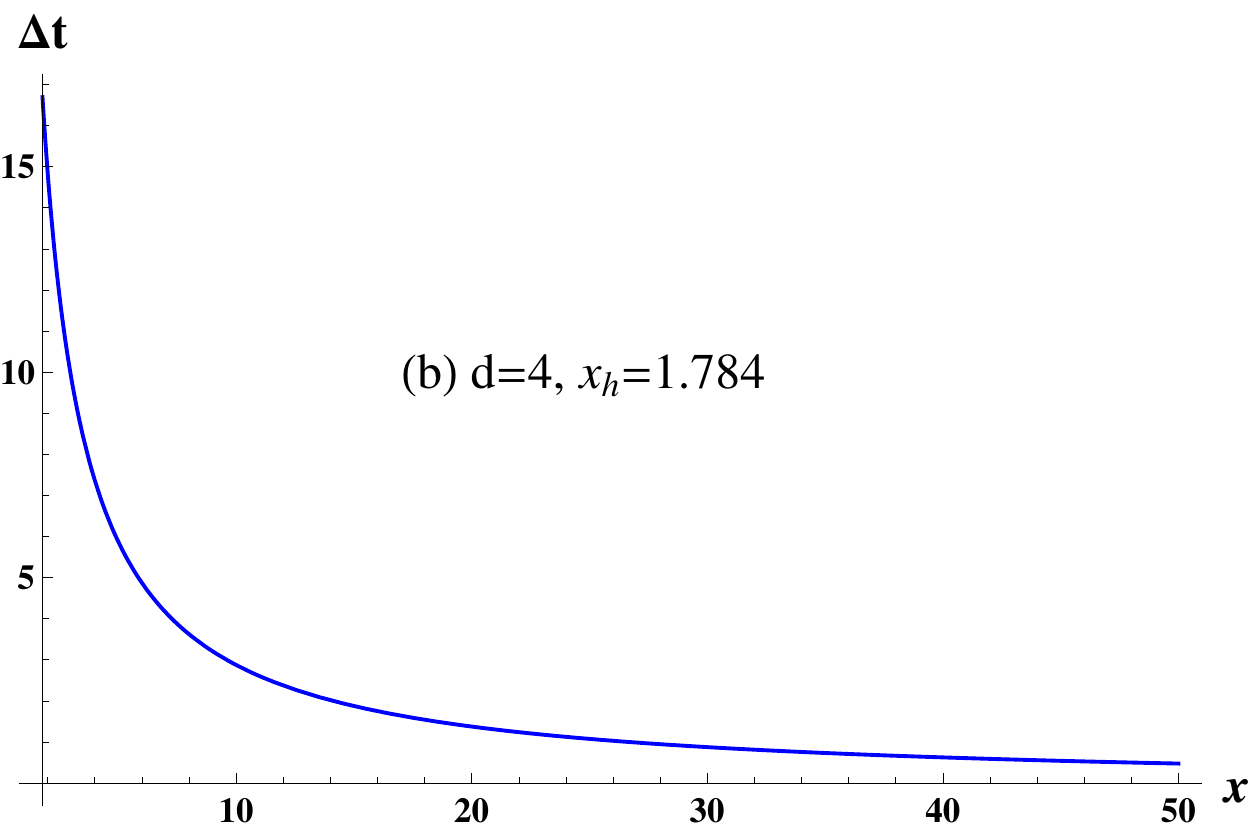}}
{\includegraphics[width=.5\textwidth]{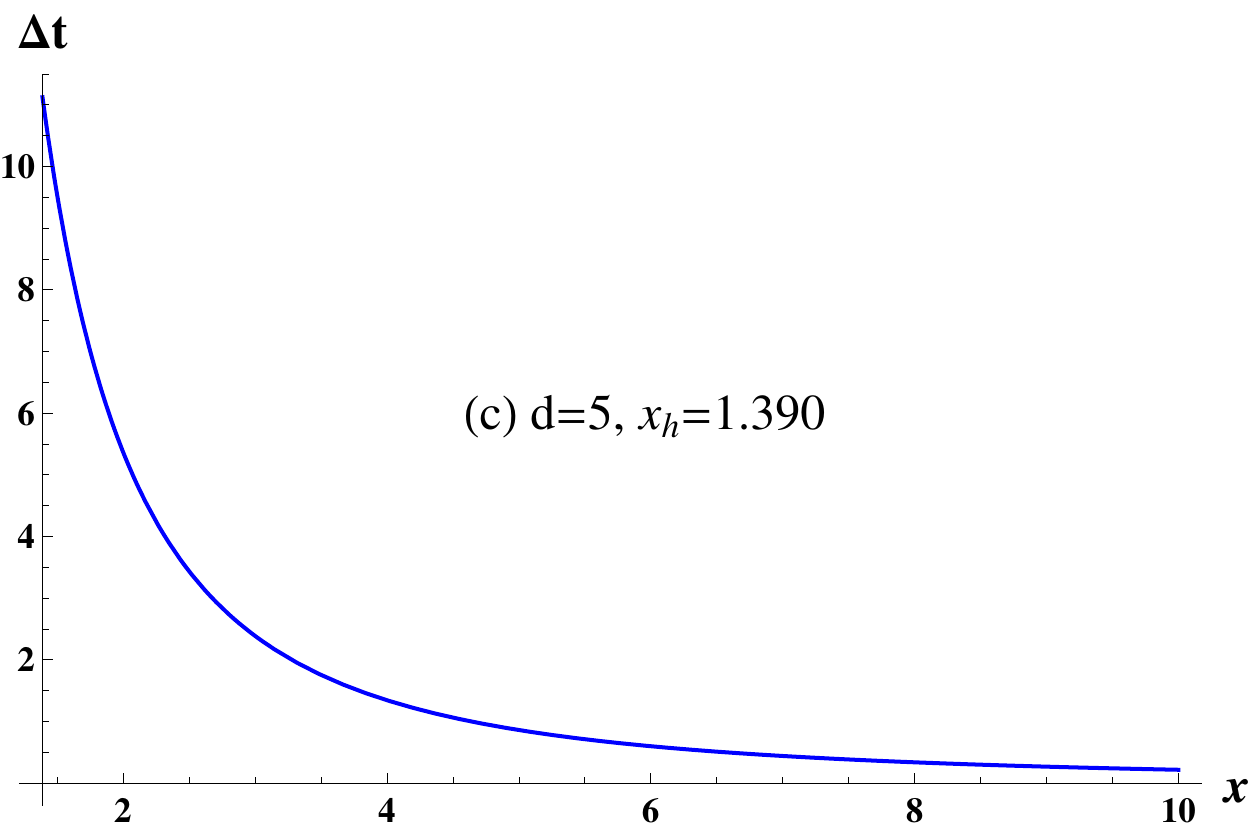}}
{\includegraphics[width=.5\textwidth]{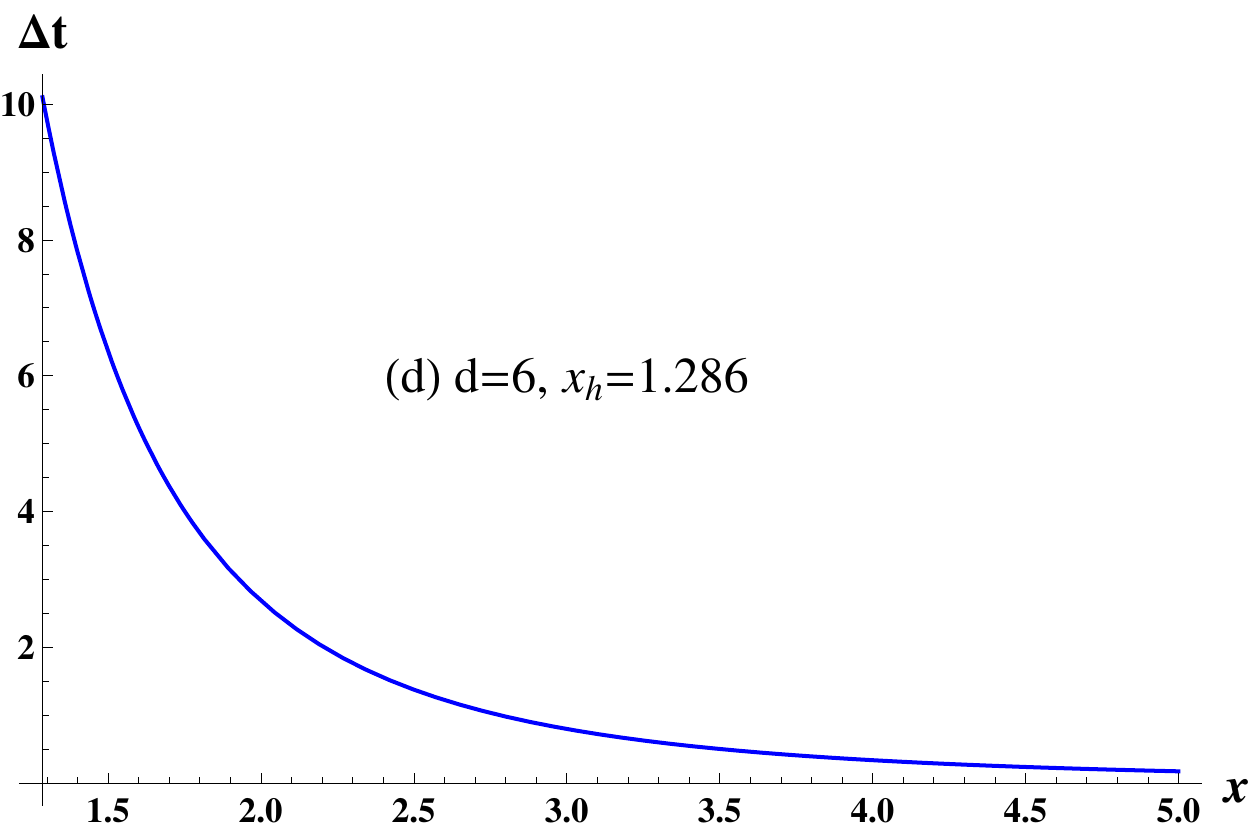}}
	    \caption{\small{Excess time delay for a higher dimensional Schwarzschild black hole with $\alpha=1.5$, $\eta=2.5$, $\delta=100$ and $\sigma=100$. In the case of maximum excess time delay the distances $\delta$ and $\sigma$ are very greater than the location of horizon.}}
\label{f4}
\end{figure}

 In order to better understand the implication of this excess delay, we have plotted (\ref{tds}) as a function of different distances from the event horizon for $d=3,4,5$, and $6$ in Figs.~(\ref{f4}). It is observed from the figures that it takes shorter time for a radar signal to travel to the planet and back in higher dimensional spaces compared to GR which again approves that, gravity is weaker in higher dimensions than GR. The diagrams are plotted from the event horizon in each dimension, however, their behaviour is the same and a maximum of time delay occurs when the signal passes close to the horizon.

 \section{Gravitational effects in higher dimensional non-commutative spaces}
 In a noncommutative geometry, point-like objects cannot exist, because there is no physical distance smaller than a minimal position uncertainty of the order of $\sqrt{\theta}$. This effect is implemented in spacetime through de-localization of matter which results in a regular or curvature singularity free, metric. This is exactly what is expected from the existence of a minimal length. The effects of this spreading over space is mathematically implemented by replacing position Dirac-delta function everywhere with a Gaussian distribution of minimal width $\sqrt{\theta}$ \cite{Rizzo:2006zb,Spallucci:2009zz}. Motivated by this result, we choose the mass density of a smeared, static, spherically symmetric source as
\be\label{gd} \rho_{M}(r)=\frac{M}{(4\pi \theta)^{d/2}} \exp{\left(-\frac{r^2}{4\theta}\right)}\,,\ee
i.e. the particle mass $M$ is diffused throughout a region of linear size  $\sqrt{\theta}$. It is generally assumed that $\sqrt{\theta}$ is close to the Planck length. However, one can define the line element and Einstein’s equation with de-localized matter sources which give regular metrics\cite{Nicolini:2005vd,Rizzo:2006zb}.

The particle-like $(d+1)$-dimensional solution of Einstein's equation with this source is described by the metric (\ref{Sch1}) \cite{Rizzo:2006zb,Spallucci:2009zz} with
\be \label{ncss}  B^{NC}(r)=1-\frac{\mu(d)  [\sqrt{\theta }]^{d-2}}{r^{d-2}} \,\gamma\left(\frac{d}{2},\frac{r^2}{4 \theta}\right),\ee
where $NC$ refers to the non-commutative space and the dimensionless parameter $\mu(d)$ is defined as follows
\be \label{mu} \mu(d)=\frac{8 M L^{d-3} }{(d-1) [\pi \theta]^{(d-2)/2}}.\ee
The Euler lower Gamma function $\gamma(a/b,z)$ is defined by
\be
\gamma(a/b,x)\equiv \int^{x}_{0}  \, e^{-t} \,t^{a/b}\,\frac{dt}{t},
\ee
and the physical mass of the solution is given by integrating the minimal
spread Gaussian profile (\ref{gd})
\be \label{mass} M_{\theta}=A_{(d-1)}\int r^2 \rho_{M}(r) dr\,.\ee

\begin{figure}[H]
\centering
\includegraphics[width=12cm,height=7cm]{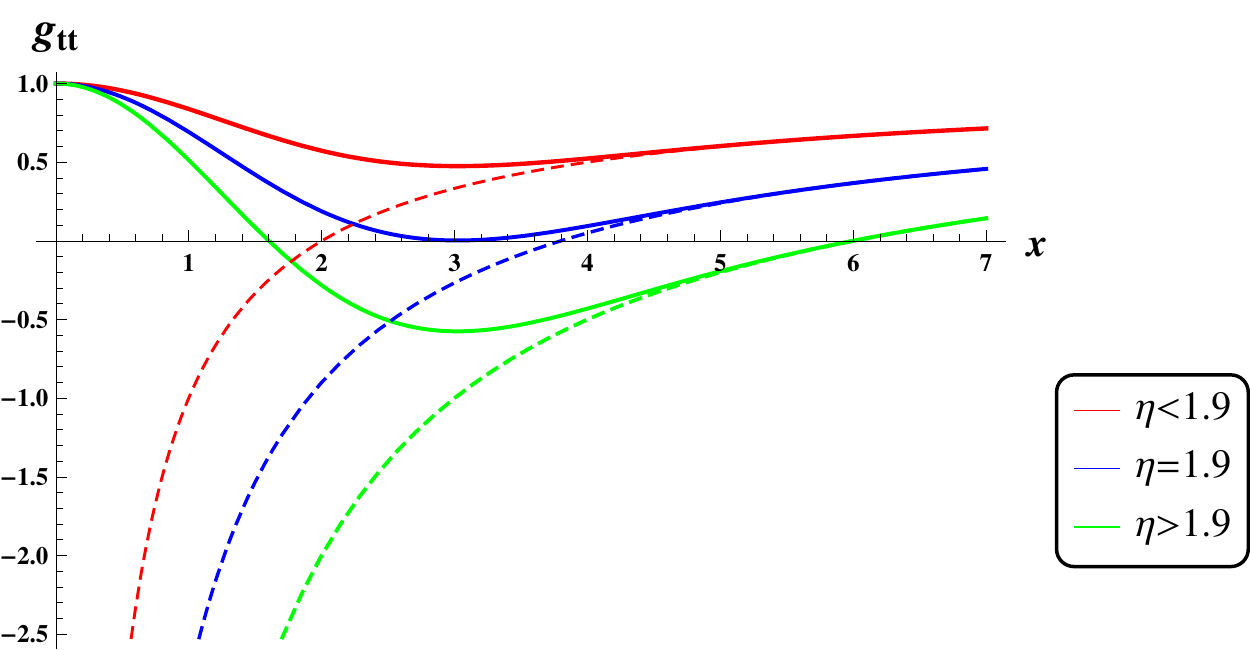}
\caption{{$g_{tt}$ in terms of $x$ for various values of $\frac{M}{\sqrt{\theta}}$. Intercepts on the horizontal axis give radii of the event horizon(s). The dashed curves show the commutative case for different values of mass $M$.}}
\label{mss}
\end{figure}

For an observer at large distances, $\frac{r}{\sqrt{\theta}}\rightarrow \infty$ or $\frac{\sqrt{\theta}}{r}\rightarrow 0$, this smeared density looks like a small sphere of matter with radius about $\sqrt{\theta}$, so it assures that the metric to be Schwarzschild.
In contradiction to the usual Schwarzschild black hole in GR, which has a single horizon, in (3+1)-dimensional non-commutative space we have different possibilities. An important and interesting question in a non-commutative background is: what is the condition to have a black hole with one (extremal) or two horizons. For a Schwarzschild metric in (3+1)-dimensional space, it is shown in \cite{Nicolini:2005vd} that:
\begin{itemize}
\item For $\eta=\frac{M}{\sqrt{\theta}}<1.9$ there is no horizon for (\ref{ncss}) with $d=3$ shown by the red curve in Fig.~(\ref{mss})
\item For $\eta=\frac{M}{\sqrt{\theta}}=1.9$ there is a degenerate horizon (extremal black hole) in $x=\frac{r}{\sqrt{\theta}}=3$ shown by the blue curve in Fig.~(\ref{mss}). This mass is called the $minimal\, mass\,, M=M_0=1.9\sqrt{\theta}$ \cite{Nicolini:2005vd}, which represents the final state of a black hole at the end of Hawking evaporation process.
\item For $\eta=\frac{M}{\sqrt{\theta}}>1.9$ there are two distinct horizons shown by green curve in Fig.~(\ref{mss}). By increasing $M$, i.e., for $M >> M_0$, the inner horizon shrinks to zero, while the outer horizon approaches the Schwarzschild value $r = 2M$.
\end{itemize}

In fact, the location of event the horizon for a Schwarzschild-like black hole is determined by the equation $B(r)\!=\!0$ in our convention, but one cannot exactly solve this equation for $r_h$ in non-commutative geometry. So, we should solve it numerically for different values of mass parameter $M$ as illustrated in Fig. (\ref{mss}) for $d=3$ dimensions. This is also true for the parameter $\mu(d)$ in non-commutative extra dimensions where one can say that it has the same role as the black hole mass $M$ for Schwarzschild black hole in metric function (\ref{ncss}). It is natural to ask the same question in the context of non-commutative extra dimensions as at what values of  $\mu(d)$can a black hole exist (having at least one degenerate horizon). We denote this value of $\mu(d)$ by $\mu_0 (d)$, and the relevant values are calculated and summarized in  Tab.~(\ref{tab1}).

\begin{figure}[H]
\centering
\includegraphics[width=14cm,height=8cm]{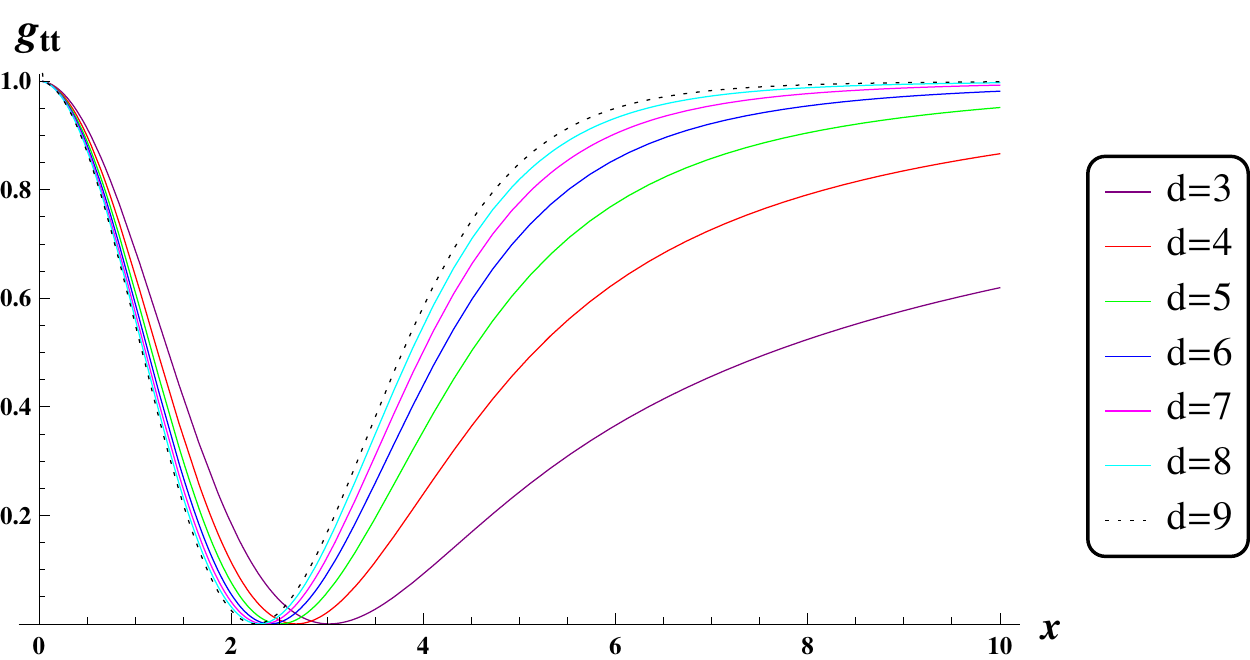}
\caption{{Time component of the extremal Schwarzschild metric in different dimensions vs. $x=\frac{r}{\sqrt{\theta}}$.}}
\label{newmet}
\end{figure}

 \begin{table}[H] 
 \caption[]{Values for $\mu_0 (d)$(Minimum of $\mu(d)$ to form a black hole) in different dimensions.}
\centering
\begin{tabular}{|c|c|c|c|c|c|c|c|}\hline
 $d$&$3$&$4$&$5$&$6$&$7$&$8$&$9$  \\ \hline
 $\mu_0 (d)$ & $4.29714$&$13.40368$&$36.813$&$94.11858$&$229.84576$&$543.7545$&$1256.8274$       \\ [0.2ex]\hline
\end{tabular}
\label{tab1}
\end{table}
To check the existence of horizons and their radii we have plotted (\ref{ncss}) as a function of $\frac{r}{\sqrt{\theta}}$ using the values shown in Tab.~(\ref{tab1}) which is depicted in Fig.~(\ref{newmet}). As expected, the black holes do exist in all dimensions i.e., there is one degenerate horizon in each dimension.

\subsection{Gravitational measurements}
Now, we are ready to study the redshift and deflection of light around a higher- dimensional non-commutative Schwarzschild-like geometry.

\begin{figure}[H]
\centering
\includegraphics[width=14cm,height=8cm]{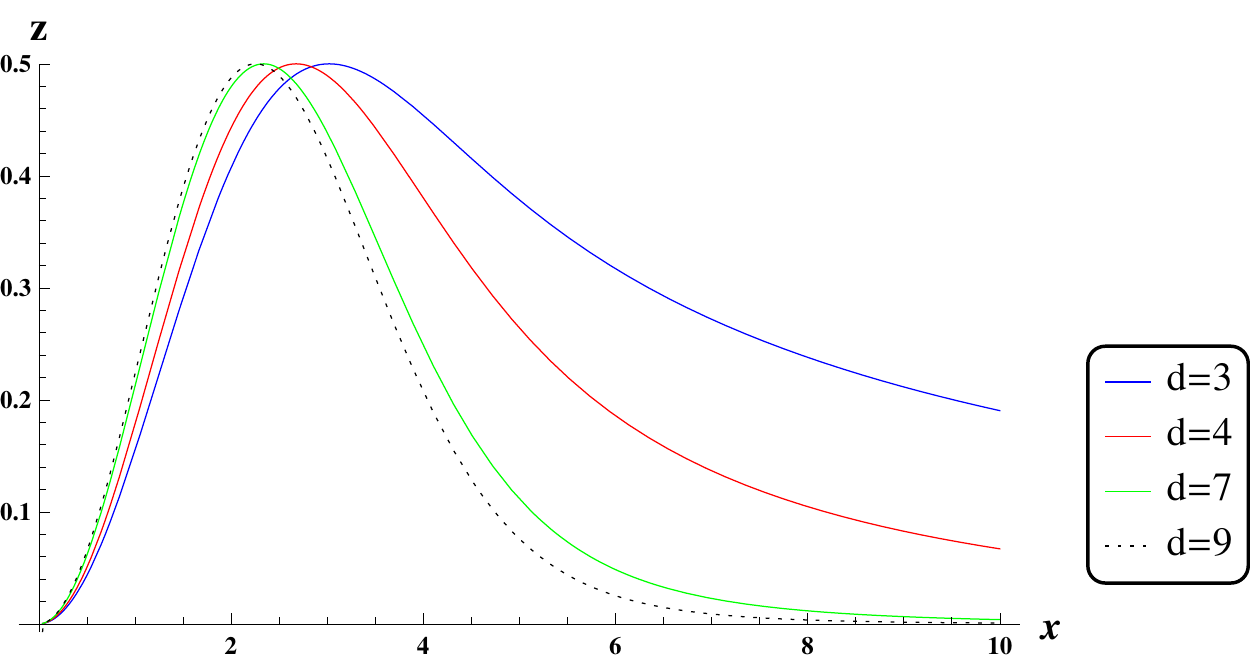}
\caption{{Redshift of a higher dimensional NC Schwarzschild black hole.}}
\label{srnred}
\end{figure}

{\bf{\emph{Redshift:}}}
In the context of non-commutative geometry in CCS approach, the redshift function is obtained by inserting  (\ref{ncss}) in (\ref{reds1}) and doing necessary calculation,  so the maximum redshift measured by an asymptotic observer, $r_2\rightarrow \infty$, is given by
\be\label{ncsred} z^{NC}=\left[1-\frac{\mu(d)  [\sqrt{\theta }]^{d-2}}{r_1^{d-2}} \,\gamma\left(\frac{d}{2},\frac{r_1^2}{4 \theta}\right)\right]^{-1/2} \!-\!1,\ee
where in the limit $\frac{\sqrt{\theta}}{r_1}\rightarrow 0$, it leads to (\ref{reds2}) for the higher dimensional commutative Schwarzschild solution.
Using (\ref{mu}) and Tab.~(\ref{tab1}), we have plotted the redshift function calculated by (\ref{ncsred}) for different spatial dimensions in Fig.~(\ref{srnred}) in terms of dimensionless radial coordinate $x=\frac{r_1}{\sqrt{\theta}}$. As expected, far away from the gravitational system, all the curves tend to zero and there is no shift in the light wavelength just as in the commutative spaces.

We can also see from Fig.~(\ref{srnred}) that in contrast to the commutative space, there is a maximum of redshift (peak) in each dimension. However, the values of the peaks are the same for all dimensions, i.e., it does not depend on the dimension and by increasing the dimension of spacetime, the peak occurs in smaller $x$. For instance, there is a regular peak at $x=3$ for extremal limit $\eta=1.9$ in four-dimensions \cite{Karimabadi:2018sev}. It has been shown in \cite{Nicolini:2005vd} that there exists a similar finite maximum temperature at $r_h=3\sqrt{\theta}$ that the black hole can reach before cooling down to absolute zero which states that there is no curvature singularity at the origin and the geometry is regular there. It is also observed for higher dimensions that by increasing $x=\frac{r_1}{\sqrt{\theta}}$, the redshift decreases more rapidly than four-dimensional spacetime, which shows that gravity becomes weaker in higher dimensions.

\begin{figure}[H]
\centering
\includegraphics[width=14cm,height=8cm]{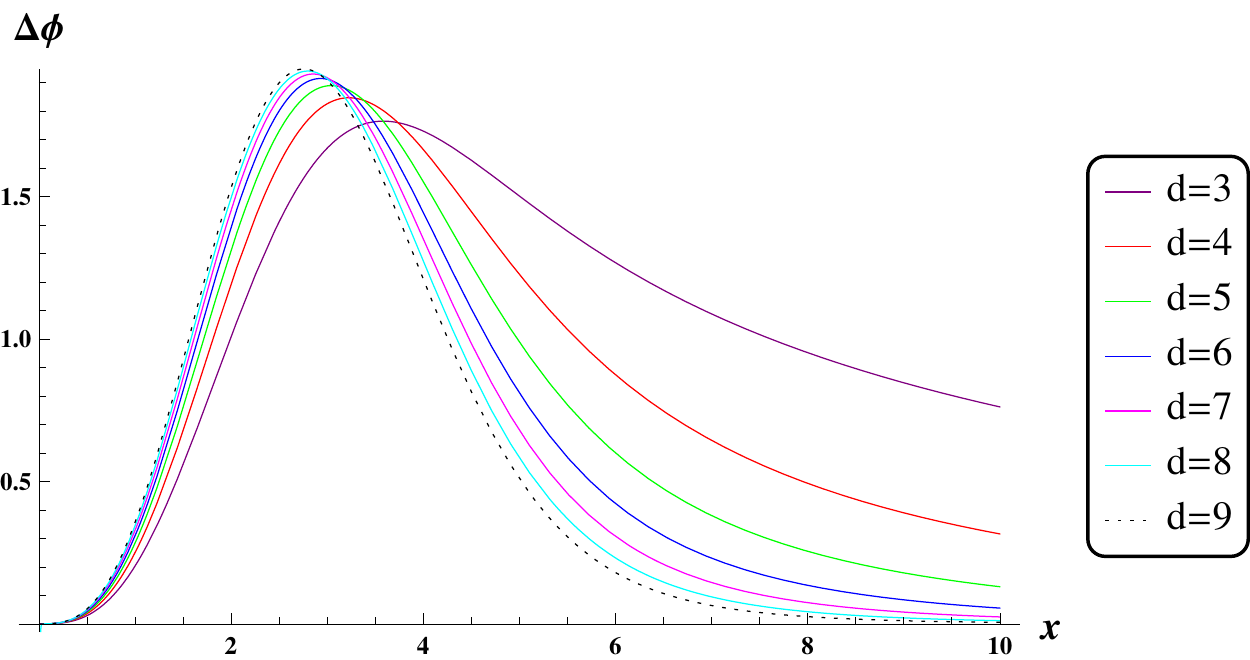}
\caption{{Gravitational deflection close to a non-commutative Schwarzschild black hole for $\eta=1.9$ and $\alpha=3$.}}
\label{f8}
\end{figure}

{\bf{\emph{Deflection of light:}}}
The amount of deflection of light when passing close to a higher dimensional Schwarzschild black hole in a non-commutative space is calculated by inserting the metric (\ref{ncss})  in the relation (\ref{def}), so we have
\bea\label{defsh1} \Delta\phi&\!\!\!\!=\!\!\!\!&-\pi+2\int_{r_{\circ}}^{\infty} dr \,\Big[\frac{1}{r \sqrt{\frac{r^2}{r_{\circ}^2}-1}}+\frac{4Mr\,  L^{d-3}}{(d-1)\, \pi^{\frac{d-2}{2}}  r_{\circ}^{d}\,\left(\frac{r^2}{r_{\circ}^2}-1\right){}^{3/2}}\,\gamma\left(\frac{d}{2},\frac{r^2}{4 \theta}\right)\nn\\
&\!\!\!\!-\!\!\!\!&\frac{4M  r^{3-d} L^{d-3}}{(d-1)\, \pi^{\frac{d-2}{2}} r_{\circ}^{2} \,\left(\frac{r^2}{r_{\circ}^2}-1\right){}^{3/2}}\,\gamma\left(\frac{d}{2},\frac{r^2}{4 \theta}\right)+\frac{4M  r^{1-d} L^{d-3}}{(d-1)\, \pi^{\frac{d-2}{2}} \,\left(\frac{r^2}{r_{\circ}^2}-1\right){}^{1/2}}\,\gamma\left(\frac{d}{2},\frac{r^2}{4 \theta}\right)\Big] ,\eea
and after integration the result is as follows
\be\label{defsh2} \Delta\phi^{NC}=\frac{8M  L^{d-3}}{(d-1)\, \pi^{\frac{d-3}{2}}  r_{\circ}^{d-2}} \,\gamma\left(\frac{d+1}{2},\frac{r_{\circ}^2}{4 \theta}\right)\,, \ee
where in the limit $\frac{\sqrt{\theta}}{r}\rightarrow 0$, it gives the predicted deflection as denoted by (\ref{Schdef}). There are again two points that can be inferred from Fig.~(\ref{f8}); $i$) There is a regular peak for the value of deflection in each spacetime dimension at the degenerate horizon. $ii$) The maximum value reduces by decreasing the spacetime dimensions and also the peak of the deflection takes place in a smaller $x$ as we increase the number of dimensions. $iii$) As we get away from the horizon, the value of deflection goes faster to zero by increasing spacetime dimensions and it is because there are more channels available for gravity to decay in them.

\subsection{Determination of lower bounds}

{\bf{\emph{Lower bound on the size of extra dimensions (L):}}}
One can impose the condition $L>r_{\circ}$ ($\frac{L}{\sqrt\theta}>\frac{r_{\circ}}{\sqrt{\theta}}$) to ensure that L is larger than the event horizon radius and so is an observable quantity, see also \cite{Scardigli:2003kr}. The values of $\frac{r_{\circ}}{\sqrt{\theta}}$ are extractable from Fig.~(\ref{newmet}). Also, using the fact that the length of $\sqrt{\theta}$ is less than $7\times10^{-19}$ \cite{Karimabadi:2018sev}, one can obtain a lower bound on the length of extra dimensions L as  provided in Tab.~(\ref{tab3}).

 \begin{table}[H] 
 \caption[]{Lower bound on length of extra dimensions.}
\centering
\begin{tabular}{|c|c|c|}\hline
 $d$ & ${r_{\circ}}/{\sqrt{\theta}} $ & $ L(m)$ \\  \hline

 4 & 2.68&$1.87\times10^{-18}$ \\
 5 & 2.51&$1.1\times10^{-18}$ \\
 6 & 2.41&$9.38\times10^{-19}$ \\
 7 & 2.34&$8.65\times10^{-19}$\\
 8 & 2.29&$8.26\times10^{-19}$ \\
 9 & 2.26&$8.01\times10^{-19}$ \\[0.5ex]\hline
\end{tabular}
\label{tab3}
\end{table}

{\bf{\emph{Lower bound on the mass of the black hole (M):}}}
As mentioned earlier the metric (\ref{ncss}) for special values of $\mu(d)$, which are listed in Tab.~(\ref{tab1}), corresponds to extremal black holes in different extra dimensions. So the condition
\be\label{A} \mu(d)\geq \mu_0 (d)\,, \ee
ensures that we have black holes with two distinct horizons (the greater sign) or with one degenerate horizon (the equal sign). Using (\ref{A}) and the results of Tab.~(\ref{tab3}), one can provide a lower bound on the mass to form a black hole in higher dimensional spacetime the results of which are summarized in Tab.~(\ref{tab4}). These values are close to the order of the mass of the primordial micro black holes created in the early Universe which may have survived (have not evaporated) until the current epoch \cite{Liddle:1998nt,Sato-Polito:2019hws}. 
 \begin{table}[H] 
 \caption[]{Lower bound on the mass of the black hole (M).}
\centering
\begin{tabular}{|c|c|}\hline
 $d$ & $M(kg)$ \\  \hline
 3 & $1.79\times10^{9}$ \\
 4 & $5.54\times10^{9}$ \\
 5 & $3.83\times10^{10} $\\
 6 & $2.26\times10^{11} $\\
 7 & $1.21\times10^{12} $\\
 8 & $6.07\times10^{12} $\\
 9 & $2.86\times10^{13} $\\[0.5ex]\hline
\end{tabular}
\label{tab4}
\end{table}
\section{Conclusions}

In this paper, we studied the well-known tests of GR for higher dimensional commutative and non-commutative Schwarzschild black holes. We obtained expressions for the gravitational redshift, deflection, and time delay around black holes. The results show that the amounts of these quantities will diminish when we study higher dimensional black holes. In this regard, as depicted in Figs.~(\ref{f2}), (\ref{f3}) and (\ref{f4}), by increasing the dimensions of spacetime in commutative case, the effects of gravity become weaker than GR, what is consistent with the fact that the gravitational effects propagate into the extra dimensions, or that gravity gets diluted in the large volume of the extra dimensions \cite{ArkaniHamed:1998rs}.

On the other hand, in a non-commutative geometry which is based on the CCS formalism, we observed that the existence of a Schwarzschild black hole with a degenerate horizon (extremal black hole) tightly depends on $M$, $L$ and $\sqrt{\theta}$, i.e. the mass of the black hole, the size of the extra dimensions and the non-commutative length scale, respectively. It has been shown in Fig.~(\ref{newmet}) and Tab.~(\ref{tab1}) that for a definite higher dimensional non-commutative geometry, i.e., for given values of $L$ and $\sqrt\theta$, by increasing the number of dimensions, we need more mass to generate extremal Schwarzschild black holes. In spite of GR, where the redshift factor does not have a finite value, in the non-commutative case, there is a finite extremum value in which light might shift to the red wavelength.

Very interesting and important in higher dimensional research, we have also obtained a minimum mass needed to form a black hole in each dimension and a lower bound on the size of the extra dimensions. Our results also confirm the previous studies that hold gravity becomes weaker in extra dimensions. Also, it would be of interest to investigate the effects of the charged solutions in each sector, that is, for both commutative (higher dimensional RN black holes) and non-commutative spaces \cite{Spallucci:2009zz}. In conclusion, if there exist extra dimensions of space in nature, as it seems to emerge from different theories and arguments, then the implications should appear in GR gravitational measurements such as those treated in this work.

\section*{Acknowledgment}
The authors would like to thank Asghar Moulavinafchi for his careful reading of the manuscript and for his valuable comments which helped us improve the English language and grammar in our paper.

\end{document}